\documentclass[twocolumn,superscriptaddress,pre]{revtex4}
\usepackage{epsfig}
\usepackage{amssymb,amsfonts,amsmath}
\usepackage[pdftex]{pstricks}

\include{epsf}

\def\(({\left(}
\def\)){\right)}                       
\def\[[{\left[}
\def\]]{\right]}
\def\e{{\rm e}}

\newcommand{\ben}{\begin{enumerate}}
\newcommand{\een}{\end{enumerate}}

\newcommand{\<}{\langle}
\renewcommand{\>}{\rangle}

\newcommand{\beq}{\begin{equation}}
\newcommand{\eeq}{\end{equation}}
\newcommand{\bea}{\begin{eqnarray}}
\newcommand{\eea}{\end{eqnarray}}

\renewcommand{\phi}{\varphi}
\newcommand{\SI}[1]{App.~#1 \cite{SI}}

\usepackage{amsmath}
\usepackage{amssymb}
\usepackage{amsthm}
\usepackage{amsfonts}

\begin{document}
\title{Physical limit to concentration sensing in a changing environment}
\author{Thierry Mora}
\thanks{Corresponding author: thierry.mora@ens.fr}
\affiliation{Laboratoire de physique de \'Ecole normale sup\'erieure
  (PSL University),
  CNRS, Sorbonne University, Universit\'e de Paris, 24 rue Lhomond,
  75005 Paris, France}
\author{Ilya Nemenman}
\affiliation{Department of Physics, Department of Biology, and
  Initiative in Theory and Modeling of Living Systems, Emory University, Atlanta, GA 30322, USA}

\linespread{1}

\begin{abstract}
Cells adapt to changing environments by
sensing ligand concentrations using specific receptors. The accuracy of
sensing is ultimately limited by the finite number of ligand molecules
bound by receptors.
Previously derived physical limits to sensing accuracy 
have assumed that the concentration was constant and ignored
its temporal fluctuations.
We formulate the problem of concentration
sensing in a strongly fluctuating environment as a non-linear
field-theoretic problem, for which we find an excellent approximate Gaussian solution.
We derive a new physical bound on the
relative error in concentration $c$ which scales as $\delta c/c \sim
(Dac\tau)^{-1/4}$ with ligand diffusivity $D$, receptor
cross-section $a$, and characteristic fluctuation time scale $\tau$, in stark contrast with the
usual Berg and Purcell bound $\delta c/c \sim (DacT)^{-1/2}$ for a
perfect receptor sensing concentration during time $T$. We show how the bound can
be achieved by a simple biochemical network downstream the receptor
that adapts the kinetics of signaling as a function of the square root
of the sensed concentration.

\end{abstract}

\maketitle

Cells must respond to extracellular signals to guide their actions in
the world. The signals typically come in the form of changing
concentrations of various molecular ligands, which are conveyed to the
cell through ligand binding to cell surface receptors. A lot of ink
has been expended on deriving the fundamental limits to the precision
with which a cell can measure the concentrations from the activity of
its receptors, constrained by the stochasticity of ligand
binding and unbinding \cite{Berg1977,Bialek2005,Kaizu2014,Aquino2016}. In
particular, it has become clear that the temporal sequence of
binding-unbinding events carries more information about the underlying
ligand concentration than just the mean receptor occupancy, typically
used in deterministic chemical kinetics models of this problem
\cite{Endres2009b}. In particular, such precise temporal information
allows cells to estimate the concentration of a cognate ligand even in a sea
of weak spurious ligands \cite{Siggia2013,Lalanne2015a,Mora2015b}, as well as to estimate
concentrations of multiple ligands from fewer receptor types
\cite{Singh2017,Singh2019}, and molecular network motifs able to perform such
complex estimation exist in the real world, even potentially taking
advantage of cross-talk between receptor-ligand pairs \cite{Carballo-Pacheco2019}.

Importantly, concentrations of ligands are worth measuring only when
they are \textit{a priori} unknown; or, in other words, if they change
with time, allowing for instance cells to adapt their behaviour
accordingly and maximize their long-term growth \cite{Kussell2005}. However, all of the preceding analyses have focused on the
regime with a clear time scale separation, where the concentration is
constant or constantly changing \cite{Mora2010a}
during the period over which it is estimated. In this article, we will fill
in this gap by calculating the accuracy with which a temporally
varying ligand concentration may be estimated from a sequence of
binding and unbinding events. This requires making assumptions about
the time scale over which significant changes of the concentration are
possible. In our formulation, the optimal sensor performs a Bayesian
computation, formalized mathematically as a stochastic field
theory. Crucially, we show how simple biochemical circuits allow one
to perform the relevant complex computations.

{\em Field theory of concentration sensing.} We associate to the ligand
concentration $c(t)$ a field $\phi(t)$ through $c(t)=c_0e^{-\phi(t)}$,
  where $c_0$ is an irrelevant reference concentration.
Ligand concentration controls the ligand-receptor binding rate
$r(t)=4Da c(t)=4Dac_0e^{-\phi(t)}\equiv r_0e^{-\phi(t)}$, where $4Da$ is the
diffusion-limited binding rate per molecule of the ligand to its
target receptor, modeled as a circle of diameter $a$ on the cell's
surface, and $D$ is the ligand diffusivity. This
binding rate can be readily generalized to $N$ receptors by using instead
$r(t)=4NDac(t)$. All our results will then hold with this additional $N$ factor. We assume that the concentration follows a geometric random walk, with
characteristic time scale $\tau$: $d\phi = \tau^{-1/2}dW$,
with $W$ a Wiener process. This choice is justified by the fact
that in many biological contexts, such as bacterial
chemotaxis, concentrations may vary over many orders of magnitude.

The probability of the
concentration temporal evolution over the time interval $(0,T)$ is given by
\beq\label{eq:prior}
P_{\rm prior}(\{\phi(t)\})=\frac{1}{Z_{\rm prior}}\exp\left[-\frac{\tau}{2}\int_0^T dt\,{\left(\frac{d\phi}{dt}\right)}^2\right].
\eeq
The receptor sees binding events at times $t_1,t_2,\ldots,t_n$, each
occuring with rate $4Dac(t_i)=r_0e^{-\phi(t_i)}$. To
simplify, let us assume that unbinding is instantaneous
(generalization to finite binding times is discussed later). The posterior distribution of the
concentration profile then follows Bayes' rule:
\begin{multline}
P(\{\phi(t)\})= \frac{P(t_1,\ldots,t_n|\{\phi(t)\})P_{\rm prior}(\{\phi(t)\})}{P(t_1,\ldots,t_n)}\\
=\frac{1}{Z}\exp\left\{-\int_0^T dt\,\left[\frac{\tau}{2}{\left(\frac{d\phi}{dt}\right)}^2+r_0e^{-\phi(t)}\right]-\sum_{i=1}^n \phi(t_i)\right\},
\label{Bayes}
\end{multline}
where $Z$ is a normalization constant independent of $\phi$. The term $r_0e^{-\phi}dt$ in the
integral corresponds the probability of not binding a ligand between
$t$ and $t+dt$ (except at
times $t_i$). The binding events at $t=t_i$ are generated by
the {\em true} temporal trace of ligand concentration,
$c^*(t)=c_0e^{-\phi^*(t)}$. In the following the true trace $\phi^*(t)$ will be
distinguished from the field $\phi$, which refers to our observation-based belief.

The one-dimensional field-theoretic problem \eqref{Bayes} is
a particular case of Bayesian
filtering \cite{Chen2003}. When collecting information from binding
events, cells do not have access to
the future and cannot use the full span $[0,T]$ of
observations to infer the concentration at time $t$. Instead, they
must infer it solely based on past observation in the interval
$[0,t]$, which distinguishes our problem from the mathematically similar
inference of a continuous probability density \cite{Bialek1996,Nemenman2002,Kinney2014,Kinney2015,Chen2018}. This inference can be performed
recursively by the rules of Bayesian sequential
forecasting, similar to the
transfer matrix technique, and also known as
the forward algorithm \cite{Chen2003}. To do this recursion, we first define:
\begin{multline}
Z(\phi,t)=\int\mathcal{D}\phi(t)\,\delta(\phi(t)-\phi) \exp\left[-\frac{\tau}{2}\int_0^t dt'\,{\left(\frac{d\phi}{dt}\right)}^2\right. \\\left.-\int_0^t dt'\,\left(r_0e^{-\phi(t')}+\phi(t')\sum_{i=1}^n \delta(t'-t_i)\right)\right].
\end{multline}
Considering past observations during the interval $[0,t]$, the posterior distribution
of $\phi$ at time $t$ reads: 
\beq
P(\phi,t)=\frac{Z(\phi,t)}{Z(t)}, \quad\textrm{with}\quad Z(t)=\int_{-\infty}^{\infty} d\phi' Z(\phi',t).
\label{PthruZ}
\eeq

When considering periods during which no binding event was observed,
we can write a recursion for $Z(\phi,t)$ between $t$ and $t+dt$. Taking
the $\delta t\to 0$ limit yields, for $t\neq t_i$ (\SI{A}):
\beq\label{eq1}
\frac{\partial P(\phi,t)}{\partial
  t}=-r_0(e^{-\phi}-\<e^{-\phi}\>)P(\phi,t)+\frac{1}{2\tau}\frac{\partial^2
  P}{\partial \phi^2},
\eeq
where $\<\cdot\>$ denotes an average over $P(\phi)$.
When a binding even does occur at time $t_i$, the posterior
distribution is updated using Bayes' rule:
\beq\label{eq2}
P(\phi,t_i^+)=\frac{e^{-\phi} P(\phi,t_i^-)}{\<e^{-\phi}\>},
\eeq
where $t_i^{\pm}$ refer to the values right before and after the
observation. The partition function $Z(t)$ can be similarly calculated
(\SI{A}) and could in principle be used to infer the correct timescale
$\tau$ by maximizing $P(\tau|\{t_1,\ldots,t_N\})\propto Z$ (\SI{C}).

{\em Gaussian solution.} Because of the $P(\phi)$ dependence
in $\<e^{-\phi}\>$, the equations for the evolution of the
posterior probability
\eqref{eq1}-\eqref{eq2} are nonlinear. However, assuming a Gaussian
Ansatz $P(\phi,t)=(2\pi\sigma(t)^2)^{-1/2}\exp[-(\phi-\hat
\phi(t))^2/2\sigma(t)^2]$, which is accurate in the limit of long
measurement times (see below), gives a closed-form solution (\SI{B}), with:
\bea
\frac{d\hat\phi}{dt}&=&\sigma^2\left[r_0e^{-\hat\phi+\sigma^2/2}-\sum_{i=1}^n
  \delta(t-t_i)\right],\label{eq:gaussian}\\
\frac{d\sigma^2}{dt}&=&\frac{1}{\tau}-\sigma^4r_0e^{-\hat\phi+\sigma^2/2}.\label{eq:gaussian2}
\eea
The maximum a posterior estimator for the concentration is then simply
given by $\hat c(t)=c_0e^{-\hat \phi(t)}$, while $\sigma(t)^2$ defines
the Bayesian uncertainty on the estimator.

\begin{figure}[t]
\begin{center}
\noindent\includegraphics[width=\linewidth]{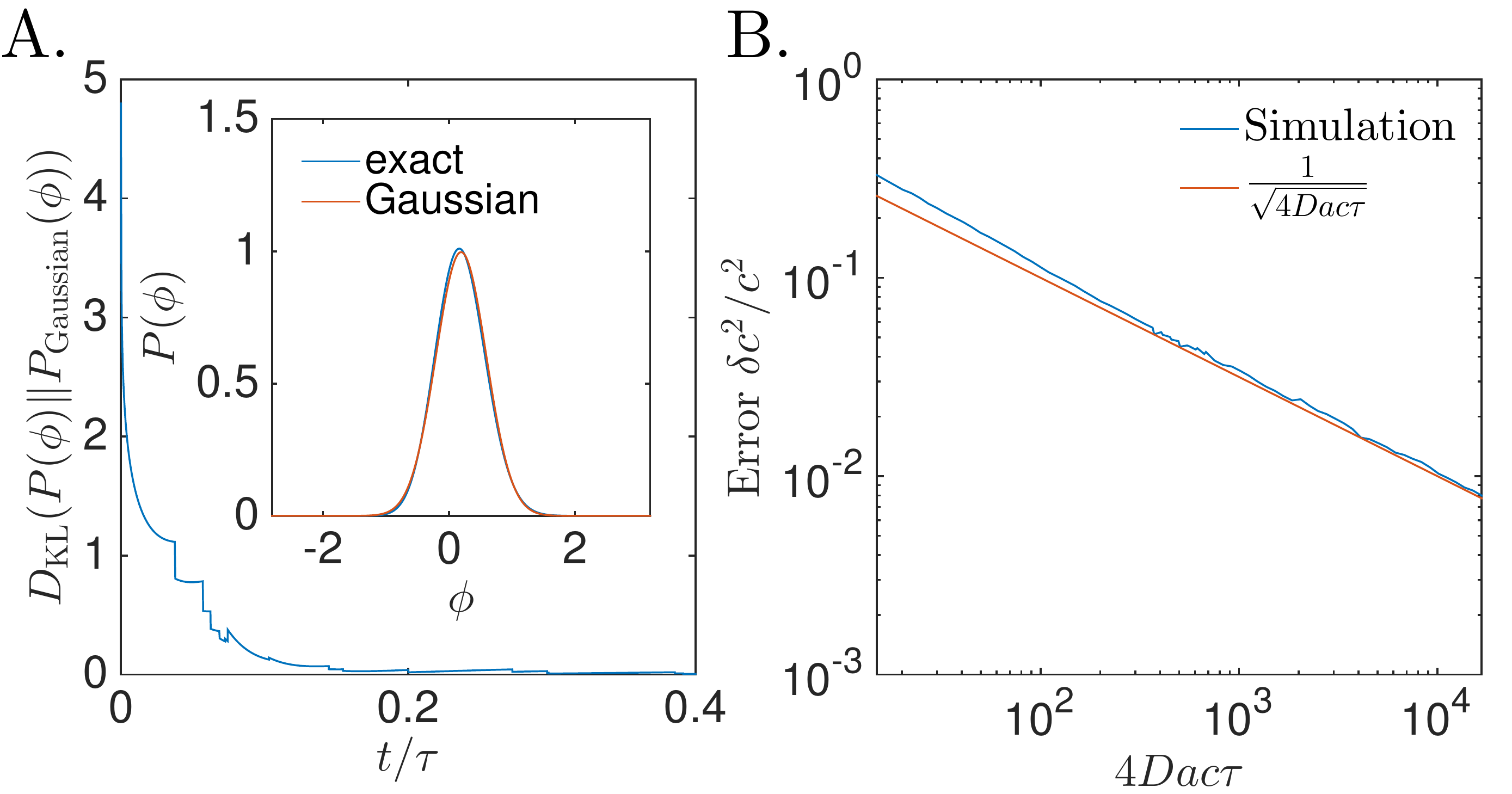}
\caption{{\bf Numerical validations of analytical results.} {\bf A.}
  The Gaussian Ansatz \eqref{eq:gaussian}-\eqref{eq:gaussian2} is
  validated by simulating the general equations for Bayesian filtering
  \eqref{eq1}-\eqref{eq2}. The numerical solution approaches the
  Gaussian solution rapidly, as indicated by the decay of the
  Kullback-Leibler divergence $D_{\rm KL}(P(\phi)\Vert P_{\rm
    Gaussian}(\phi))=\int d\phi\, P(\phi)\ln(P(\phi)/P_{\rm
    Gaussian}(\phi))$. We used $r\tau=Dac\tau=50$.
{\bf B}. Concentration sensing error as a
  function of concentration. The error estimated from simulations
  follows closely the prediction from \eqref{bprev}, which is expected
  to be valid for $4Dac\tau\gg 1$.
\label{fig1}
}
\end{center}
\end{figure}

To check the validity of the Gaussian solution, we simulated
\eqref{eq1}-\eqref{eq2} numerically, starting from a uniform
distribution ($P(\phi,0)=1/2$ for $\phi\in[-1,1]$ and $0$ otherwise), with
$r_0\tau=50$ and a true $\phi^*(t)$ starting at $\phi^*(0)=0$. The numerical
solution quickly approaches the Gaussian solution given by
\eqref{eq:gaussian}-\eqref{eq:gaussian2} starting with $\hat
\phi(0)=\<\phi\>_{t=0}$ and
$\sigma(0)^2=\mathrm{Var}(\phi)_{t=0}$. The Kullback-Leibler
divergence between the numerical and analytical solutions falls rapidly (Fig.~\ref{fig1}A) and the numerical
solution approaches the predicted Gaussian very closely
(Fig.~\ref{fig1}A, inset). Thus, the Gaussian solution provides an excellent
approximation.

{\em Error estimate.} To study the typical behaviour of
\eqref{eq:gaussian}-\eqref{eq:gaussian2}, we now assume that the rate
of binding events is large compared to the rate of change of the
concentration, $4Dac\tau=r\tau\gg 1$. This regime is the biologically
relevant one: to sense concentration, cells need to record many
binding events over the time scale on which the concentration
fluctuates. In that limit the estimator $\hat \phi$ is close to the
true value $\phi^*$, and the Bayesian uncertainty $\sigma^2$ is small,
allowing for two simplifications. First, \eqref{eq:gaussian2} relaxes over time
scale $r(t)^{-1}$ to a quasi-steady state value $\sigma^2\approx 1/\sqrt{r_0
e^{-\hat \phi}\tau}\ll 1$. Second, we can make a small noise
approximation for binding events: over some time interval $\Delta t$,
with $r^*(t)^{-1}\ll \Delta
t\ll \tau$, the number of binding events has both mean and variance
equal to $r^*(t)\Delta t$, allowing us to replace discrete jumps in
\eqref{eq:gaussian} by:
\beq\label{smallnoise}
d\left(\sum_{i=1}^n \delta(t-t_i)\right)\approx  r_0e^{-\phi^*}dt + (r_0e^{-\phi^*})^{1/2}dW',
\eeq
where $W'$ is a Wiener process. As a result, the estimator $\hat \phi$
tracks the true value $\phi^*$ according to:
\beq
d\hat \phi \approx (r_0 e^{-\hat \phi}/\tau)^{1/2} (\phi^*-\hat \phi)+\tau^{-1/2}dW',
\eeq
where we have expanded at first order in $\hat\phi-\phi^*$. In the
general case, the true field may evolve according to a different
characteristic time scale, $\tau^*$, than the one assumed by the
Bayesian filter, $\tau$, so that $d\phi^*=(\tau^*)^{-1/2}dW$. The
estimation error $\epsilon=\hat\phi-\phi^*$ then evolves according to:
\beq
d\epsilon=-(r/\tau)^{1/2}\epsilon\,dt
+\tau^{-1/2}dW'-(\tau^*)^{-1/2}dW.
\eeq

Intrigingy, the noises $dW'$ and $dW$ have very different
interpretations, one being due to the random arrival of
binding events, and the other to the geometric diffusion of the
concentration. Yet they come in the same form in this equation.
Relying again on the assumption that $r\tau\gg 1$, we get an estimate of the
error:
\beq\label{eq:epsilon}
\<\epsilon^2\>=\frac{1}{2\sqrt{r}}\left(\frac{1}{\sqrt{\tau}}+\frac{\sqrt{\tau}}{\tau^*}\right),
\eeq
which has a minimum as a function of $\tau$, reached for the true
value of the characteristic fluctuation time $\tau=\tau^*$:
\beq\label{bprev}
\frac{\<(\hat c-c^*)^2\>}{c^2}\approx \<\epsilon^2\>=\frac{1}{\sqrt{r\tau }}=\frac{1}{\sqrt{4Dac\tau}}.
\eeq
This error is equal to the Bayesian uncertainty
$\sigma^2=1/\sqrt{r_0\tau e^{-\hat \phi}}\approx 1/\sqrt{4Dac\tau}$
and is consistent with the error found using the
saddle-point approximation in the related problem of probability
density estimate \cite{Bialek1996}.

We checked the validity of our small-noise approximation by comparing
the prediction from \eqref{eq:epsilon} with the results of a numerical
simulation of \eqref{eq:gaussian}-\eqref{eq:gaussian2}, in which we averaged the error $\<(\hat c-c^*)^2\>$ as a
function of $c$ for many realization of the process. The agreement is
found to be
excellent, and gets better as $r\tau=4Dac\tau$ becomes larger (Fig.~\ref{fig1}B).

The error in \eqref{bprev} sets a fundamental physical limit on any
concentration sensing device, biological or
artificial, in a concentration profile that follows a geometric
random walk.
This bound is radically different
from that obtained by Berg and Purcell for the concentration
sensing error by a single
receptor integrating binding events over time $T$ \cite{Berg1977,Endres2009b}:
\beq
\frac{\delta c^2}{c^2}=\frac{1}{4DacT}
\eeq
(in the limit where binding events are short so that the
receptor is always free).

The major difference is that Berg and Purcell, as well as most of the
literature on concentration sensing, assume that the
sensed concentration does not change with time. Our result can be
reconciled with Berg and Purcell by defining an effective measurement
time $T\sim\sqrt{\tau/4Dac}$ --- the geometric mean between the mean time between
binding events and the time
scale of variation. This $T$ realizes the optimal tradeoff between
the requirement to integrate over many binding events, $T\gg 1/(4Dac)$,
but over a relatively constant concentration, $T\ll \tau$ \cite{Kolmogorov1939}.

\begin{figure}[t]
\begin{center}
\noindent\includegraphics[width=.29\linewidth]{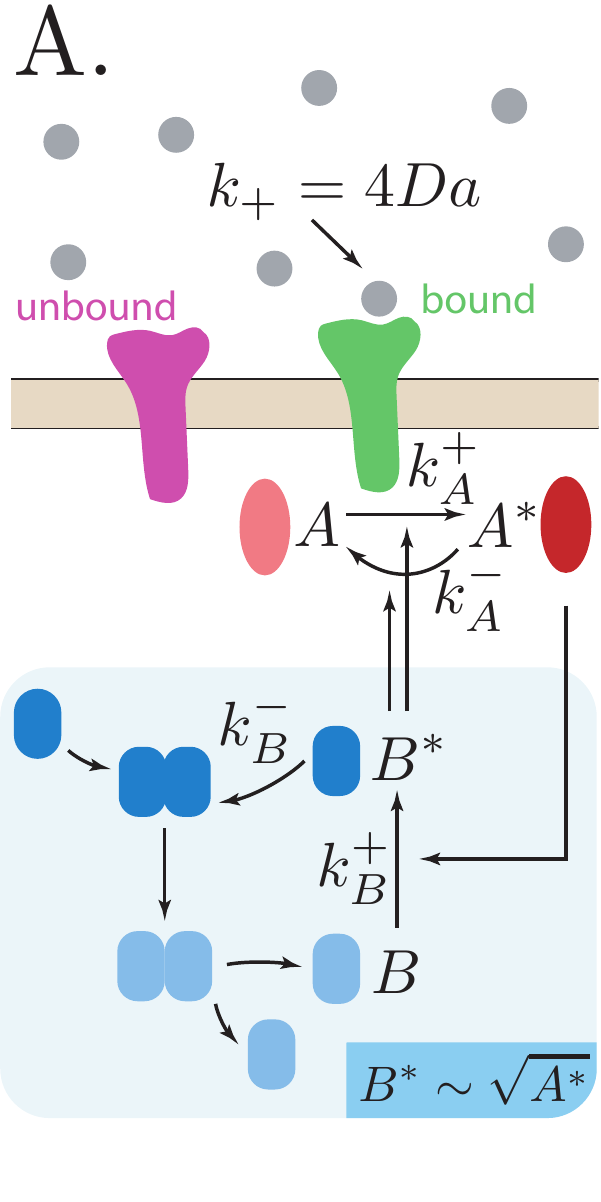}
\includegraphics[width=.69\linewidth]{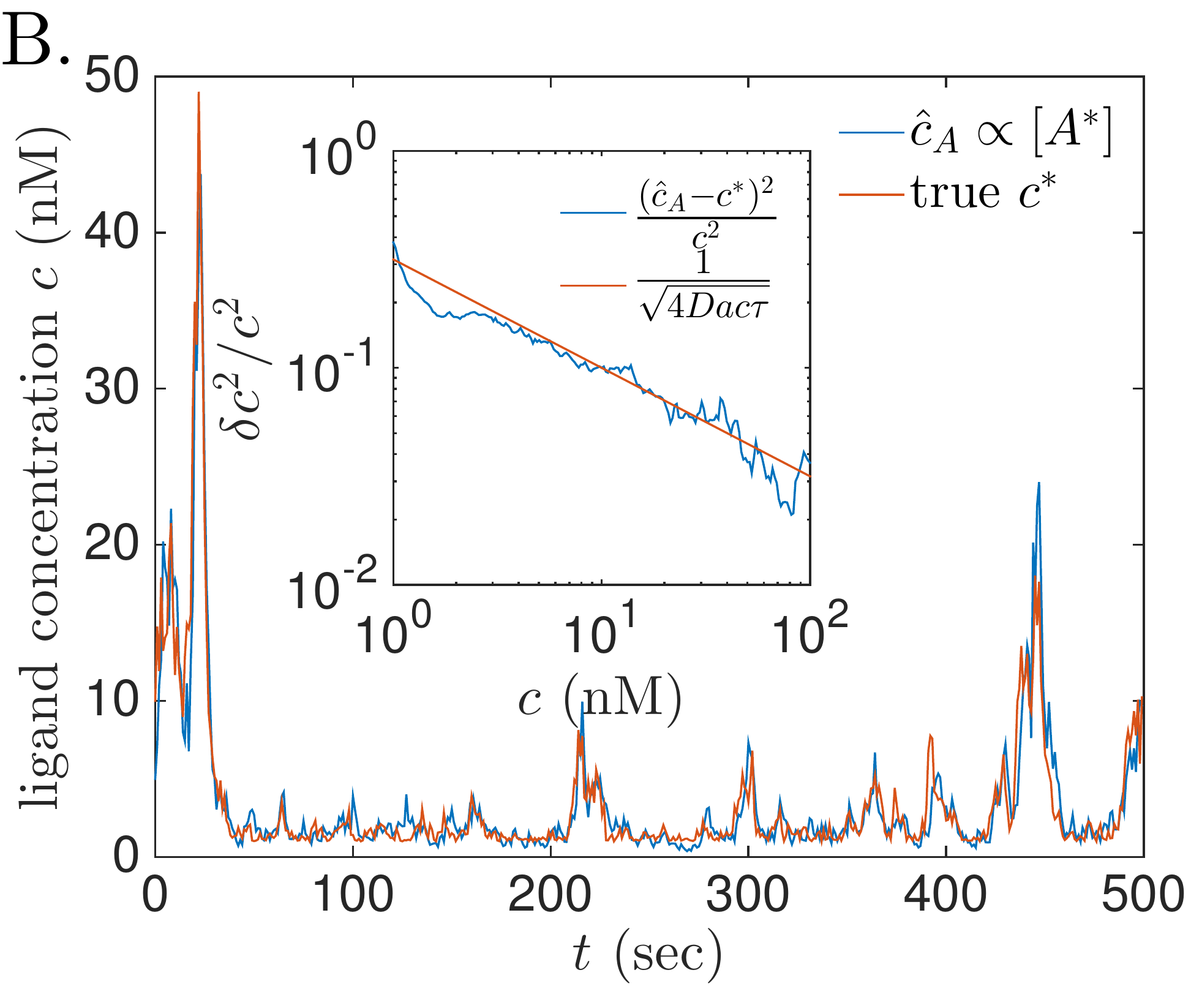}
\caption{{\bf Performance of adaptive biochemical network in
    fluctuating ligand concentration.} {\rm A.} Schematic of the
  biochemical network implementing optimal Bayesian filtering. The
  receptor-induced activation of the readout molecule $A^*$, as well as its deactivation  are regulated by a second molecule $B^*$, which is made to
  scale like $\sqrt{A^*}$ using a mechanism of deactivation by
  dimerisation (shaded box). {\bf B.} Simulation of the network
  readout $c_A(t)\propto A^*(t)$ in response to stochastic binding events in
  a fluctuating  concentration field $c^*(t)$. The relative estimation error
  $\<(\hat c_A-c^*)^2\>/c^2$  behaves according to the theoretical bound
  $1/\sqrt{4Dac\tau}$ (inset).
\label{fig2}
}
\end{center}
\end{figure}

{\em Plausible biological implementation.}
Can cells implement the optimal Bayesian filtering scheme and reach
the bound set by \eqref{bprev}? To gain intuition, it is useful to
rewrite \eqref{eq:gaussian}-\eqref{eq:gaussian2} in term of the
concentration estimator $\hat c$, in the limit $4Dac\tau\gg 1$ where
$\sigma^2$ can be eliminated:
\beq\label{eq:r}
\frac{d\hat c}{dt}=\sqrt{4Da\hat c/\tau}\left(\frac{1}{4Da}\sum_{i=1}^n\delta(t-t_i)-\hat c\right).
\eeq
Each binding event should lead to an increment of $\hat c$, followed
by a continuous, exponential decay, with a rate
given by $T^{-1}=\sqrt{4Da\hat c/\tau}$.

This scheme can be implemented by a simple biochemical network
schematized in Fig.~\ref{fig2}A.
The concentration readout $\hat c_{A}$ may be
represented by the ``active'' (for instance phosphorylated) form $A^*$ of a chemical species. Binding events
cause the receptor to activate $A$ into $A^*$, which gets subsequently deactivated.
Both the activation and deactivation of $A$ are catylized by a second chemical species in its active form,
$B^*$. Thus, upon a binding event, the concentration of active $A^*$
is increased by:
\beq\label{eq:DeltaA}
\Delta [A^*] = k_A^+ [A] [B^*],
\eeq
and it decays between binding events according to:
\beq\label{eq:decayA}
\frac{d[A^*]}{dt}=-k_A^- [B^*] [A^*],
\eeq
where $k_A^\pm$ are biochemical parameters.

To implement \eqref{eq:r}, the concentration of $B^*$ must be
controled by the square root of $A^*$. This dependence can be achieved
by assuming that $B$ is
activated into $B^*$ through the catalityc activity of $A^*$, and that
$B^*$ gets deactivated cooperatively as a dimer:
\beq\label{eq:B}
\frac{d[B^*]}{dt}=k^+_B [B]  [A^*] - k^-_B[B^*]^2,
\eeq
where $k_B^\pm$ are biochemical reaction rates.

Assuming that the kinetics of $B$ are fast compared to $A$,
we obtain $B^*=(Bk^+_B/ k^-_B)^{1/2} \sqrt{A^*}$ and
\beq
\label{astar}
\frac{d[A^*]}{dt}=\alpha \sqrt{[A^*]}\\\left(\beta\sum_{i=1}\delta(t-t_i)- [A^*]\right).
\eeq
with $\alpha=k_A^-({[B]k^+_B/ k^-_B})^{1/2}$ and $\beta=(k_A^+[A]/k_A^-)$.
If $A$ and $B$ are in excess, and thus approximately constant, then
this biochemical network exactly implements (\ref{eq:r}), with $
4Da \hat c_A\equiv k_A^-[A^*]/k_A^+[A]$, and $\tau=\tau_{\rm
  net}\equiv 1/(\alpha^2\beta)=k^-_B/
  (k^+_Bk_A^+k_A^- [A][B])$.

Interestingly, the amount of inactive ($\approx$ total) $B$ controls
the time scale of concentration fluctuations, and could be tuned
through gene regulation to adapt to different speeds of environmental
fluctuations. A biochemical network might be able to find the optimal
$\tau$ and then adjust $[B]$ accordingly by empirically measuring the
fold-change of $r(t)$ (which can be done by biochemical networks,
see e.g. \cite{Goentoro2009}) but with a delay, $\<r(t+\Delta t)/r(t)\>=e^{\Delta
  t/2\tau}$, and then inverting the relationship to extract $\tau$.

We tested the performance of the biochemical network for sensing
concentration by simulating \eqref{eq:DeltaA}-\eqref{eq:B} with a
fluctuating ligand concentration $c(t)$ with characteristic
time scale $\tau$. For concreteness, we set $c^*(0)=10$nM,
$\tau^*=10$s, $k_A^+[A]=0.01$, $k_A^-=k_B^+=k_B^+=1\mu
{\rm   M}^{-1}{\rm s}^{-1}$ and $[B]=10\mu$M, so that
$\tau_{\rm net}=\tau^*$. Fig.~\ref{fig2}B shows the network estimate $\hat c_A(t)$
along with the true value $c^*(t)$. The empirical error $\<(\hat
c_A-c^*)^2\>$ as a function of $c^*$ averaged over $10^4$s
(Fig.~\ref{fig2}B, inset), again shows an excellent agreement with the
theoretical bound $1/\sqrt{4Dac\tau}$.

{\em Discussion.} For the sake of clarity our analysis made
simplifying assumptions which can be easily relaxed.
Our proposed biochemical implementation assumed a constant burst of
activity following each binding event, consistent with the optimal estimation strategy. However, in
real receptors, stochasticity in the bound time is known to double the variance
in the estimate
\cite{Endres2009b} (\SI{D}). Treating this effect simply adds a factor
$\sqrt{2}$ in the noise term of \eqref{smallnoise} as well as in
\eqref{bprev}, $\<\delta
c^2\>/c^2\approx 1/\sqrt{2Dac\tau}$.
We also ignored periods during which
the receptor was bound. During that time the receptor is blind to the
external world, and the posterior evolves according to the prior:
$\partial_t P=(1/2\tau) \partial^2_\phi P$, $\partial_t
\hat\phi=0$ and $\partial_t\sigma^2=1/\tau$. In our results, these
``down times'' renormalize
the effective observation time by the fraction
of time the receptor is free, $p_{\rm free}=(1+4Dacu)^{-1}$,
where $u$ is the average bound time, $\<\delta
c^2\>/c^2\approx 1/\sqrt{4Dacp_{\rm free}\tau}$ (\SI{D}). Combining the two
effects (stochasticity in bound time and receptor availability) would
yield $\<\delta
c^2\>/c^2\approx 1/\sqrt{2Dacp_{\rm free}\tau}$.

The field theory of \eqref{Bayes} is mathematically similar to the problem
of estimating a density function from a small sample set with a smoothing
prior \cite{Bialek1996,Nemenman2002,Kinney2014,Chen2018}. The main
difference lies in the domain of observations. In density estimation the
whole function $\{\phi(t)\}_{t\in[0,T]}$ is infered together on the
whole domain of $t$, while sensors can only learn from past
observations, {\em i.e.} the $t'<t$ half-plane. However, our solution
can easily be generalized to deal with the entire time domain using
the forward-backward algorithm (\SI{E}). Eqs.~\eqref{eq1}-\eqref{eq2} and
\eqref{eq:gaussian}-\eqref{eq:gaussian2} can be solved both forward
(from $0$ to $t$) and backward (from $T$ to $t$, with time reversal) in time, giving
$P_{\to}(\phi)$, $\hat \phi_\to$, $\sigma^2_\to$ for the forward
solution (the one treated in this article), and
$P_{\leftarrow}(\phi)$, $\hat \phi_\leftarrow$,
$\sigma^2_\leftarrow$ for the backward solution. The Bayesian
posterior at any given time is then given by
$\propto P_{\to}(\phi) P_{\leftarrow}(\phi)$, of mean
$(\sigma^2_{\leftarrow}\hat\phi_\to+\sigma^2_{\to}\hat\phi_\leftarrow)/(\sigma^2_{\leftarrow}+\sigma^2_{\to})$
and variance
$\sigma^2_{\leftarrow}\sigma^2_{\to}/(\sigma^2_{\leftarrow}+\sigma^2_{\to})$
in the Gaussian approximation.
While this situation is not relevant for
concentration sensing, our general solution should be
applicable to problems of density estimation. The saddle-point
approximation usually made in that context \cite{Bialek1996,Nemenman2002,Kinney2014} is expected to work in the same
limit as our Gaussian Ansatz; however, recent work has emphasized the importance of non-Gaussian
fluctuations for small datasets
\cite{Chen2018}.

The biological implementation we propose is speculative. An
interesting direction would be to identify square-root or similar control of
receptor signaling in real biological systems, and interpret them in terms of optimal Bayesian
filtering. Signaling pathways dealing with concentration changes over
several orders of magnitude, such as bacterial chemotaxis, typically
use adaptation mechanisms to increase the dynamic range of sensing
\cite{Lazova2011a}---a feature that is absent from our approach as
we neglect noise in the signaling output. Combining adaptation
design with ideas from Bayesian estimation could help us gain insight
into the fundamental bounds and resource allocation tradeoffs
that limit biological information processing.

{\em Acknowledgments.} The authors would like to thank the Casa
Matem\'atica Oaxaca from the Banff International Research Station
where this work was initiated. TM was partially supported by Agence National
pour la Recherche (ANR) grant No. ANR-17-ERC2-0025-01
``IRREVERSIBLE'' and IN by NSF Grants No.\ PHY-1410978 and IOS-1822677.

\onecolumngrid

\appendix

\section{Field theory for concentration sensing} 
We first recall the problem outlined in the main text for self-consistency.
Receptor binding happens with rate $r(t)=4Dac(t)=r_0e^{-\phi(t)}$. The
field $\phi(t)$ follows a random walk with characteristic time
$\tau^*$: 
\beq
d\phi(t)=\tau^{-1/2}dW.
\eeq
In the following $\phi^*$ will refer  to the actual realization of the
random walk, with characteristic time $\tau^*$, and $\phi$ will refer
to the guess made based on the observation of binding events, while
$\tau$ denotes the assumes characteristic time scale.

The probability of the time trace of $\phi$ in the absence of any
observation is given by
\beq\label{eq:priorsi}
P_{\rm prior}(\{\phi(t)\})=\frac{1}{Z_{\rm prior}}\exp\left[-\frac{\tau}{2}\int_0^T dt\,{\left(\frac{d\phi}{dt}\right)}^2\right],
\eeq
with $Z_{\rm prior} = (2\pi dt/\tau)^{T/2dt}$, where $dt$ is an
infinitesimal discretization scale.

During each interval $[t,t+dt]$ without a binding event, the likelihood reads
$e^{-dt r(t)}=e^{-dt r_0e^{-\phi(t)}}$. A binding event in 
interval $[t,t+dt]$ has likelihood $dt r_0e^{-\phi(t)}$. Thus the posterior
probability thus reads:

\beq
P(\{\phi(t)\})= \frac{P(t_1,\ldots,t_n|\{\phi(t)\})P_{\rm prior}(\{\phi(t)\})}{P(t_1,\ldots,t_n)}
=\frac{1}{Z}\exp\left\{-\int_0^T dt\,\left[\frac{\tau}{2}{\left(\frac{d\phi}{dt}\right)}^2+r_0e^{-\phi(t)}\right]-\sum_{i=1}^n \phi(t_i)\right\},
\label{Bayessi}
\eeq

We define:
\beq
Z(\phi,t)=\int\mathcal{D}\phi\,\delta(\phi(t)-\phi) \exp\left\{\int_0^t
  dt'\,\left[-\frac{\tau}{2}{\left(\frac{d\phi}{dt}\right)}^2-r_0e^{-\phi(t')}-\phi(t')\sum_{i=1}^n
    \delta(t'-t_i)\right]\right\}.
\eeq
The marginal of $\phi$ at time $t$, $P(\phi,t|\{t_1,\ldots,t_{n'}\})$,
where $n'$ is the last binding event before $t$, is then: 
\beq
P(\phi,t)=\frac{Z(\phi,t)}{Z(t)}, \qquad Z(t)=\int d\phi' Z(\phi',t),
\eeq
and $Z=Z(T)$.

The partial partition function $Z(\phi,t)$ can be computed
recursively. Let us start with the case where no binding occurs
between $t$ and $t+dt$. Then:
\beq
Z(\phi,t+dt)=\int d\phi'\,Z(\phi',t)\exp\left[-\frac{\tau(\phi-\phi')^2}{2dt}-dt r_0e^{-\phi}\right].
\eeq
Equivalently,
\beq
P(\phi,t+dt)=\frac{1}{z(t)}\int d\phi'\,P(\phi',t)\exp\left[-\frac{\tau(\phi-\phi')^2}{2dt}-dt r_0e^{-\phi}\right],
\eeq
where
\beq
z(t)=\int
d\phi\,d\phi'\,P(\phi',t)\exp\left[-\frac{\tau(\phi-\phi')^2}{2dt}-dt\,
  r_0e^{-\phi}\right]
\eeq
is a normalization constant, which can be used to calculate $Z(t)$
recursively: $Z(t+dt)=Z(t)z(t)$.
Let us now take the limit $dt\to 0$. The Gaussian integral becomes
infinitely peaked. Expanding
$P(\phi',t)=P(\phi,t)+\partial P(\phi,t)/\partial
\phi (\phi'-\phi)+ \partial^2P(\phi,t)/\partial \phi^2
(\phi'-\phi)^2/2$, we obtain:
\beq
\frac{\partial P(\phi,t)}{\partial
  t}=-r_0(e^{-\phi}-\<\e^{-\phi}\>)P(\phi,t)+\frac{1}{2\tau}\frac{\partial^2
  P}{\partial \phi^2},
\eeq
where $\<f(\phi)\>=\int d\phi\, P(\phi,t) f(\phi)$, and
\beq
z(t)=(1-dt\,\<e^{-\phi}\>)\sqrt{2\pi dt/\tau}.
\eeq
Defining $Z(t)= (2\pi dt/\tau)^{t/2dt}\tilde Z(t)$, we get:
\beq
\frac{\tilde Z(t+dt)}{\tilde Z(t)}=(1-dt\,\<e^{-\phi}\>),
\quad\textrm{or}\quad \frac{d\ln\tilde Z(t)}{dt}=-\<e^{-\phi}\>.
\eeq
Now assume that there is a binding event between $t=t_i$ and $t_i+dt$. Then $P(\phi,t)$
is discontinuous at $t_i$ and terms of order $dt$ can be ignored:
\beq
P(\phi,t_i+dt)=\frac{e^{-\phi} P(\phi,t_i)}{\<e^{-\phi}\>},
\eeq
and $\tilde Z(t_i+dt)/\tilde Z(t_i) =\<e^{-\phi}\>$.

In summary, the evolution equation for $P$ reads:
\beq\label{eq:FP}
\frac{\partial P(\phi,t)}{\partial
  t}=\frac{1}{2\tau}\frac{\partial^2
  P}{\partial \phi^2}-r_0(e^{-\phi}-\<\e^{-\phi}\>)P(\phi,t) +
\left(\frac{e^{-\phi}}{\<e^{-\phi}\>}-1\right) P(\phi,t) \sum_{i=1}^n \delta(t-t_i),
\eeq
and the partition function is given by
\beq\label{eq:Z}
Z(t)= (2\pi dt/\tau)^{t/2dt}\tilde Z(t), \quad\textrm{with}\quad
\frac{\partial\ln\tilde Z}{\partial
  t}=-r_0\<e^{-\phi}\>+\ln\<e^{-\phi}\>
\sum_{i=1}^n \delta(t-t_i).
\eeq

\section{Gaussian approximation}
Because of the term $\<e^{-\phi}\>$, \eqref{eq:FP} is non-linear and
cannot be solved analytically. However, if we assume that $P(\phi,t)$
is Gaussian,
\beq
P(\phi,t)=\frac{1}{\sqrt{2\pi
    \sigma^2(t)}}\exp\left[-\frac{(\phi-\hat\phi(t))^2}{2\sigma(t)^2}\right],
\eeq
then closed equations can be obtained for the mean $\hat\phi(t)$ and
variance $\sigma^2(t)$:
\bea
\frac{d\hat\phi}{dt}&=&\frac{d\<\phi\>}{dt}=\frac{1}{2\tau}\int
d\phi \phi \frac{\partial^2
  P}{\partial \phi^2}-r_0(\<\phi e^{-\phi}\>-\<\phi\>\<\e^{-\phi}\>)
+
\left(\frac{\<\phi e^{-\phi}\>}{\<e^{-\phi}\>}-\<\phi\>\right) \sum_{i=1}^n
\delta(t-t_i)\nonumber
\\&=&
\sigma^2\left[r_0e^{-\hat\phi+\sigma^2/2}-\sum_{i=1}^n
  \delta(t-t_i)\right],\label{eq:gsi1}\\
\frac{d\sigma^2}{dt}&=&\frac{d(\<\phi^2\>-\<\phi\>^2)}{dt}=\frac{1}{2\tau}\int
d\phi \phi^2 \frac{\partial^2
  P}{\partial \phi^2}
-r_0(\<\phi^2 e^{-\phi}\>-\<\phi^2\>\<\e^{-\phi}\>)
+
\left(\frac{\<\phi^2 e^{-\phi}\>}{\<e^{-\phi}\>}-\<\phi^2\>\right) \sum_{i=1}^n
\delta(t-t_i)-\frac{d\<\phi\>^2}{dt}\nonumber\\
&=&
\frac{1}{\tau}-\sigma^4r_0e^{-\hat\phi+\sigma^2/2},\label{eq:gsi2}\\
\frac{d\ln \tilde Z}{dt}&=&-r_0e^{-\hat\phi+\sigma^2/2}-(\hat\phi-\sigma^2/2) \sum_{i=1}^n
  \delta(t-t_i),\label{eq:logZ}
\eea
where we have used the Gaussian integral rules: $\<e^{-\phi}\>=e^{-\hat\phi+\sigma^2/2}$, $\<\phi
e^{-\phi}\>=(\hat\phi-\sigma^2)e^{-\hat\phi+\sigma^2/2}$, $\<\phi^2
e^{-\phi}\>=(\sigma^2+(\hat\phi-\sigma^2)^2)e^{-\hat\phi+\sigma^2/2}$,
and $\<\phi^2\>=\hat\phi^2+\sigma^2$, and we have used integration by
parts to calculate the integrals.

\section{Partition function and time scale inference}
The most likely timescale $\tau$ can be inferred from the observations
as well by using Bayes's rule again:
\beq
P(\tau)\propto \int \mathcal{D}\phi(t)\,
P(t_1,\ldots,t_n|\{\phi(t)\})\,P_{\rm prior}(\{\phi(t)\}|\tau)\,P_{\rm prior}(\tau)=\frac{Z(\tau)P_{\rm prior}(\tau)}{Z_{\rm prior}(\tau)}={\tilde Z(\tau)}P_{\rm prior}(\tau),
\eeq
where we have used $Z_{\rm prior}(\tau)=(2\pi dt/\tau)^{T/2dt}$.

We can calculate $\tilde Z$ from the Gaussian approximation \eqref{eq:logZ}:
\beq\label{c2}
\log \tilde Z \approx -\int_0^T dt\, \hat r(t)e^{\sigma^2(t)/2} + \sum_{i=1}^n(\ln \hat r(t_i)+\sigma^2(t_i)/2).
\eeq
This expression looks like the log-likelihood of a sequence of
binding events, up to the $\sigma^2$ corretions. Bear in mind that $\hat r(t)$ is the estimated rate, not the true one $r^*(t)$. We have $\hat r(t)= r^*(t)e^{-\epsilon(t)}$, where we $\epsilon=\hat\phi- \phi^*$. Expanding in $\epsilon$ and $\sigma^2$, we obtain:
\beq
\log \tilde Z=-\int_0^T dt\, r^*(t) + \sum_{i=1}^n\ln r^*(t_i) + \int_0^T dt\, \left[r^*(t)-\sum_{i=1}^n\delta(t-t_i)\right](\epsilon(t)-\sigma(t)^2/2)  - \int_0^T dt\,r^*(t)\frac{\epsilon^2(t)}{2}.
\eeq
Both $\epsilon(t)$ and $\sum_i \delta(t-t_i)-r^*$ are
stochastic processes of mean 0. They 
are also uncorrelated with each other, so that the third term is sub-linear
in $T$. The last term, which scales with $T$, thus dominates the
$\tau$-dependent part of the likelihood. It is maximized for minimum
mean squared error, that is for $\tau=\tau^*$, as shown in the main
text. At large $T$, the $\tilde Z(\tau)$ term exponentially dominates
the prior $P_{\rm prior}(\tau)$, so that $P(\tau)$ is peaked around the maximum of $\tilde Z(\tau)$.

Eq. \eqref{c2} is an example of the usual bias-variance
tradeoff (where ``bias'' refers to errors made from overfitting the
data, and ``variance'' to errors due to limited data). At small $\tau$,  $\hat{r}$ changes rapidly, jumping when a
new binding happens, and then rapidly decreases. Thus  the term $\sum
\ln\hat{r}(t_i)$ increases, indicating increase in the goodness of
fit. At the same time $\sigma^2(t)$ increases, so that at small $\tau$
the integral term in \eqref{c2} becomes large and negative, exploding
exponentially for $\tau\to 0$. In contrast, for $\tau\to\infty$,
$\hat{r}(t_i)= n/T$, and $\sigma^2\to 0$, so that now the goodness
of fit is small.

Overall, there’s an optimal $\tau$ that maximizes $\tilde{Z}$. The
three terms in \eqref{c2} parallel the three terms (fluctuation
determinant, goodness of fit, and the kinetic term) in the
field-theoretic formulation of continuous probability density
estimation from samples \cite{Bialek1996,Nemenman2002}, and thus we
expect that maximizing $\tilde{Z}$ will result in an optimal $\tau$
not only when the true concentration undergoes a geometric random
walk, but also when it undergoes various anomalous walks \cite{Nemenman2002}.

\section{Bound time}
We have so far neglected the time the receptor remained bound the
ligand Let us denote by $t_{i,\rm off}$ the unbinding time following
the binding time at $t_i$. When the receptor is bound, $t_i<t\leq t_{i,\rm
  off}$, no information
can be obtained from the environment, and the evolution equation for
the posterior simply follows the diffusion law:
\beq
\frac{\partial P}{\partial t}=\frac{1}{2\tau}\frac{\partial^2
  P}{\partial \phi^2}.
\eeq
The rest of the time, $t_{i,\rm off}<t\leq t_{i}$, \eqref{eq:FP}
holds. Similarly, in the Gaussian approximation, we get
\beq
\frac{d\hat \phi}{dt}=0,\qquad\frac{d\sigma^2}{dt}=\frac{1}{\tau}
\eeq
for bound receptors, $t_i<t\leq t_{i,\rm
  off}$, and \eqref{eq:gsi1}-\eqref{eq:gsi2} for unbound receptors.
In the limit where binding and unbinding events are frequent compared
to $\tau$, $r\tau\gg 1$, we have:
\beq
\frac{d\sigma^2}{dt}=\frac{1}{\tau}-p_{\rm
  free}\sigma^4r_0e^{-\hat\phi+\sigma^2/2},
\eeq
where
\beq
p_{\rm
  free}(t)=\frac{\<t_i-t_{i-1,\rm off}\>}{\<t_i-t_{i-1}\>}=\frac{r(t)^{-1}}{r(t)^{-1}+u}=\frac{1}{1+r(t)u}=\frac{1}{1+4Dac(t)u},
\eeq
where $u=\<t_{i,\rm off}-t_i\>$ is the average bound time, and
$r(t)^{-1}=\<t_i-t_{i-1,\rm off}\>$ the average unbound time. The
uncertainty then reads:
\beq
\frac{\<\delta c^2\>}{c^2}=\frac{1}{\sqrt{rp_{\rm
    free}\tau}}=\frac{1}{\sqrt{4Dap_{\rm free}c}}.
\eeq

The time the receptor remains bound has another impact on the ability
to sense concentration. In the biochemical scheme proposed in the main text, each binding
event causes a fixed burst of activity $\delta(t-t_i)$, regardless of
the bound time. In the simplest receptors however, signaling occurs
during the time the receptor is bound, which is itself stochastic. We
can model this by replacing the Dirac delta by a random burst of
activity, $b_i\delta(t-t_i)$, with $b_i$ proportional to the bound
time, $b_i=(t_{i,\rm off}-t_i)/u$, so that $\<b_i\>=1$ and
$\mathrm{Var}(b_i)=1$, since the bound time is distributed exponentially according to $(1/u)e^{-(t_{i,\rm off}-t_i)/u}$. More generally we can consider $\<b_i\>=1$ and
$\mathrm{Var}(b_i)=CV$, where $0\leq CV\leq $ denotes the coefficient of variation. The
general base $\<b_i\>$ can be renormalized away into the biochemical
parameters. The special case $CV=0$ gives back the results of the main
text. When $CV>0$, the variance of $\int_{t_0}^{t_0+\Delta t}\sum_i b_i\delta(t-t_i)$ over an
interval of duraction $\Delta t$ instead reads:
\beq
\mathrm{Var}\left[\int_{t_0}^{t_0+\Delta t}\sum_i
  b_i\delta(t-t_i)\right]=\<b_i\>\mathrm{Var}(m)+\mathrm{Var}(b_i)\<m\>
=r^*\Delta t (1+ CV)
.
\eeq
where $n'$ is the number of binding events in the interval. As the
result, the noise $dW'$ in the main text gains a factor $\sqrt{1+CV}$,
and the error becomes:
\beq\label{eq:epsilonsi}
\<\epsilon^2\>=\frac{1}{2\sqrt{r}}\left(\frac{1+CV}{\sqrt{\tau}}+\frac{\sqrt{\tau}}{\tau^*}\right),
\eeq
and minimal error reached for $\tau=(1+CV)\tau^*$:
\beq
\frac{\<\delta c^2\>}{c^2}=\<\epsilon^2\>=\frac{\sqrt{1+CV}}{\sqrt{r\tau }}=\frac{\sqrt{1+CV}}{\sqrt{4Dac\tau}}.
\eeq

Taking into account both receptor occupancy and stochasticity in bound
times finally yields:
\beq
\frac{\<\delta
  c^2\>}{c^2}=\frac{\sqrt{1+CV}}{\sqrt{4Dap_{\rm free}c\tau}},
\eeq
or, in the case of complete stochastic unbinding, $CV=1$:
\beq
\frac{\<\delta
  c^2\>}{c^2}=\frac{1}{\sqrt{2Dap_{\rm free}c\tau}}.
\eeq

\section{Beyond concentration sensing -- using the future}
Information about future binding events can be exploited by using the
backward equation for $P_\leftarrow=P(\phi,t|\{t_{n'+1},\ldots,t_n\})$:
\beq\label{eq:back}
\frac{\partial P_{\leftarrow}(\phi,t)}{\partial
  t}=-\frac{1}{2\tau}\frac{\partial^2
  P_{\leftarrow}}{\partial \phi^2}+r_0(e^{-\phi}-\<\e^{-\phi}\>)P_{\leftarrow}(\phi,t) -
\left(\frac{e^{-\phi}}{\<e^{-\phi}\>}-1\right) P_{\leftarrow}(\phi,t)
\sum_{i=1}^n \delta(t-t_i),
\eeq
where $m$ is the last binding event before $t$.
We denote by $P_\to=P(\phi,t|\{t_1,\ldots,t_{n'}\})$ the solution of the forward equation
\eqref{eq:FP} discussed before. The distribution of $\phi$ at time $t$
is then given by:
\beq
\begin{split}
P(\phi,t|\{t_1,\ldots,t_n\})&\propto
P(\{t_1,\ldots,t_n\}|\phi,t)P_{\rm prior}(\phi,t) =
P(\{t_1,\ldots,t_{n'}\}|\phi,t)
P(\{t_{n'+1},\ldots,t_{n}\}|\phi,t) P_{\rm prior} (\phi,t)\\
&\propto  \frac{P(\phi,t|\{t_1,\ldots,t_{n'}\})
P(\phi,t|\{t_{n'+1},\ldots,t_{n}\})}{P_{\rm prior}(\phi,t)}
\propto {P_\to(\phi,t)P_\leftarrow(\phi,t)}
\end{split}
\eeq
where we have used the fact that the past and future where
conditionally independent given $\phi$ at time $t$, since the process
is Markovian, and a uniform prior. We thus have:
\beq
P(\phi,t)=\frac{P_\to(\phi,t)P_\leftarrow(\phi,t)}{\int d\phi'\,
  P_\to(\phi',t)P_\leftarrow(\phi',t)}.
\eeq
Using the Gaussian solution, and denoting $\hat\phi_\to,\sigma^2_\to$
the parameters of the forward solution, and
$\hat\phi_\leftarrow,\sigma^2_\leftarrow$ those of the backward
solution, we obtain:
\beq
P(\phi,t)=\frac{1}{\sqrt{2\pi\sigma^2}}\exp\left[-\frac{(\phi-\hat
    \phi)}{2\sigma^2}\right]
\eeq
with
\beq
\hat
\phi=\frac{\hat\phi_\to\sigma^2_\leftarrow+\hat\phi_\leftarrow\sigma^2_\to}{\sigma^2_\leftarrow+\sigma^2_\to},\qquad
\sigma^2=\frac{\sigma^2_\leftarrow\sigma^2_\to}{\sigma^2_\leftarrow+\sigma^2_\to}.
\eeq

These formulas could be used in estimates of density $r(t)$ from
sparse observations $t_i$, where $r(t)$ is interpreted as a density of
events, and $t$ is the variable whose density we want to infer.

\end{document}